\documentclass[a4paper,fleqn,usenatbib]{mnras}

\usepackage{newtxtext,newtxmath}

\usepackage[T1]{fontenc}
\usepackage{ae,aecompl}

\usepackage{graphicx}	
\usepackage[]{subfig}

\newcommand{\blue}{\color{\blue}}

\newcommand{\dd}{\mathrm{d}}

\title[Morphology of relaxed and merging galaxy clusters]{Morphology of relaxed and merging galaxy clusters. \\ Analytical models for monolithic Minkowski functionals}

\author[C. Schimd, M. Sereno]{
	C. Schimd,$^1$\thanks{E-mail: carlo.schimd@lam.fr}
	M. Sereno$^{2,3}$
	\\ 
$^{1}$ Aix  Marseille  Univ,  CNRS, CNES, LAM,  Marseille,  France \\
$^{2}$ INAF -- Osservatorio di Astrofisica e Scienza dello Spazio di Bologna, via Piero Gobetti 93/3, I-40129 Bologna, Italy\\
$^{3}$ INFN, Sezione di Bologna, viale Berti Pichat 6/2, 40127 Bologna, Italy
}

\date{Accepted 2021 January 25. Received 2021 January 19; in original form 2020 July 23}

\pubyear{2019}

\begin{document}
\label{firstpage}
\pagerange{\pageref{firstpage}--\pageref{lastpage}}
\maketitle

\begin{abstract}
Galaxy clusters exhibit a rich morphology during the early and intermediate stages of mass assembly, especially beyond their boundary. A classification scheme based on shapefinders deduced from the Minkowski functionals is examined to fully account for the morphological diversity of galaxy clusters, including relaxed and merging clusters, clusters fed by filamentary structures, and cluster-pair bridges. These configurations are conveniently treated with idealised geometric models and analytical formulae, some of which are novel. Examples from CLASH and LC$^2$ clusters and observed cluster-pair bridges are discussed.
\end{abstract}





\section{Introduction}
\label{sec:intro}

Morphology of galaxy clusters is an indicator of their state of relaxation and can be used to infer their formation history and evolution. As result of the gravitational dynamics of dark and luminous matter, relaxed galaxy clusters and their hosting dark matter haloes have a triaxial shape \citep{Limousin+2013}, with tendency to prolateness over oblateness especially in their final stage of evolution as assessed by high-resolution $N$-body simulations \citep[e.g.][]{Bett+2007,Maccio+2007,DespaliGiocoliTormen2014,Bonamigo+2015} and confirmed by X-ray, optical, Sunyaev-Zel'dovich (SZ), and weak-lensing measurements \citep{Cooray2000,DeFilippis+2005,Sereno+2006,Sereno+2018}. The persistence of this trend in the outskirts of clusters depends on their mass \citep{Prada+2006}, mass accretion rate \citep{DiemerKravtsov2014}, and assembly history \citep{Dalal+2008,FaltenbacherWhite2010, More+2016}.
 
The three-dimensional shape of these structures is normally described by the eigenvalues of the mass distribution or inertia tensors and related parameters such as sphericity, elongation, ellipticity, prolateness, and triaxiality \citep{SpringelWhiteHernquist2004}. These statistics are well-suited for dynamically evolved or poorly resolved clusters, however they cannot account for the rich morphology of unrelaxed structures or beyond the virial radius shown by high-quality imaging and spectroscopy. New instruments are opening indeed a golden age for a multi-wavelenth study of protoclusters, merging clusters and their filamentary environment at both low and high redshift. The clearest example in the local universe is the Virgo cluster with its different substructures identified using GUViCS \citep{Boselli+2014} and HyperLeda \citep{Kim+2016} data. At intermediate and high redshift, some spectacular illustration of rich structures are: the outskirts of Abell 2744 probed by XMM-Newton X-ray data \citep{Eckert+2015}; the proto-clusters revealed by Herschel-SPIRE from Planck candidates \citep{Greenslade+2018} or combining VUDS and zCOSMOS-Deep data \citep{Cucciati+2018}; the filaments bridging the cluster systems A399-A401 and A3016-A3017, detected combining Planck data with ROSAT \citep{A399A401-Planck} or Chandra \citep{A3016A3017-Chon+2019}; the gaseous and dusty bridge IRDC G333.73+0.37 \citep{Veena+2018}; the molecular filamentary structures around Centaurus, Abell S1101, and RXJ1539.5 probed by ALMA and MUSE \citep{Olivares+2019}; the multiple filaments within the SSA22 protocluster \citep{Umehata+2019}. Weak gravitational lensing analyses have been successful in detecting the dense environment and the correlated dark matter around the main cluster halo \citep{Sereno+2018NatAs}.

The increasingly large samples of haloes detected in optical \citep{ryk+al14,ogu+al18,mat+al19}, X-ray \citep{xxl_I_pie+al16}, or SZ surveys \citep{ble+al15,planck_2015_XXVII} demand for flexible and reliable indicators of morphology that can be applied to the full zoo of galaxy clusters. A number of statistics, such as halo concentration, peak-centroid shift, power ratio, axial ratio, and position angle, have been considered to quantify the degree of regularity and symmetry of these structures \citep{don+al16,lov+al17}. However, these indicators can fail for very irregular systems. A cluster progenitor experiences very different shapes during the merger history and the configuration of satellite halos and local environment dramatically changes. Major mergers can be followed by slow accretion along filaments until the cluster ends up in a relatively viralized final phase with a nearly regular and spherical shape.
We aim at finding a small set of morphological parameters that can in principle describes all the different phases of the merging accretion history.

In this paper, we propose to use the three non-trivial Minkowski functionals to fully characterise the morphology of spatial structures \citep{MeckeBuchertWagner1994,Mecke2000}. We show that very different morphologies, namely major mergers, multiple mergers, and filamentary structures can be suitably described by a single set of geometrically motivated parameters. We calculate analytical expressions for the triaxial ellipsoid, a $n$-fused balls model accounting for non-relaxed clusters undergoing merging, a spiky model with $n$ cylindrical branches radially connected to a central ball possibly accounting for filaments of matter feeding a central halo, and a dumbbell model describing axially-symmetric cluster-pair bridges (\S\ref{sec:MF}; details of the calculations reported in the Appendices). These systems are then classified using the so-called shapefinders deduced from the Minkowski functionals (\S\ref{sec:classification}). Conclusions are addressed in \S\ref{sec:conclusions}.

\vspace{-12pt}
\section{Morphology by Minkowski functionals: models}
\label{sec:MF}
The Minkowski functionals are a complete set of morphological descriptors that characterise the geometry and topology of a continuous body. In three dimensions they are its volume ($V$), surface area ($A$), integral mean curvature ($H$) and integral Gaussian curvature ($G$) of the surface, the latter being linearly related to the Euler characteristic $\chi$ that counts the number of connected components minus the number of tunnels plus the number of cavities \citep{Mecke2000}. According to a characterisation theorem, the Minkowski functionals are the only valuations invariant under rotations and translations and preserving additivity and continuity \citep{Hadwiger1957}. These properties along with the Steiner formula allow the calculation of $V_0\equiv V$, $V_1\equiv A/6$, and $V_2\equiv H/3\pi$, the fourth functional $V_3\equiv \chi = 1$ being trivial for isolated bodies with no tunnels and cavities as here.

\vspace{-12pt}
\subsection{Ellipsoidal model: relaxed clusters}
\label{subsec:MF:ellipsoid}

Nearly viriliazed clusters can be conveniently described as ellipsoidal haloes.
For a triaxial ellipsoid $\mathcal{E}$ with principal semi-axes $a\geqslant b\geqslant c$, with $a$ defining the polar axis and $(b,c)$ the equatorial plane, namely with $q\equiv b/a$ and $s\equiv c/a$ respectively the intermediate-to-major and minor-to-major axis ratio, the non-trivial Minkowski functionals are
\begin{subequations}\label{eq:ellipsoid}
\begin{eqnarray}
V_0^\mathcal{E} &=& \frac{4\pi}{3}a^3qs, \\
V_1^\mathcal{E} &=& \frac{\pi}{3} a^2s^2\left[1+\frac{q}{e} F(\varphi,m)+\frac{eq}{s^2}
E(\varphi,m)\right], \\
V_2^\mathcal{E} &=& \frac{aqs}{3\pi}(I_1+I_2),
\end{eqnarray}
\end{subequations}
in which $e=\sqrt{1-s^2}$, $m=e^{-1}[1-(s/q)^2]$, $F(\varphi,m)$ and $E(\varphi,m)$ are elliptic integrals of first and second kind with $\sin\varphi=e$ \citep{abramowitz1965handbook}, and $I_{1,2}$ are dimensionless integrals that we evaluate numerically in the general case; see equations (\ref{eq:ellipsoid:curvature:Ii}) in Appendix~\ref{sec:appendix:ellipsoid}. 

Analytic limits of the previous equations exist for prolate ellipsoids of revolution about axis $a$ ($a\geqslant b=c$, so that $q=s$, $m=0$, $F=E=\arcsin e$), which could account for virialised clusters, and for oblate ellipsoids of revolution about axes $c$ ($a=b\geqslant c$, so that $q=1$, $m=1$, $F=\mathrm{arctanh}~e$, $E=e$), which could account for an intermediate stage of merging. Following the notation in \citet{Schmalzing+1999}, one has
\begin{subequations}\label{eq:ellipsoid:oblateprolate}
\begin{eqnarray}
V_0^{\mathcal{E}_*} &=& \frac{4\pi}{3}r^3\lambda, \\
V_1^{\mathcal{E}_*} &=& \frac{\pi}{3} r^2 \left[ 1+\lambda f\left(\frac{1}{\lambda}\right) \right], \\
V_2^{\mathcal{E}_*} &=& \frac{2r}{3} \left[ \lambda+g(\lambda) \right],
\end{eqnarray}
\end{subequations}
where $f(x)=(\arccos x)/\sqrt{1-x^2}$, and $\{r=as,\lambda = 1/s, g(\lambda)=f(\lambda)\}$ for prolate ellipsoids ($\mathcal{E}_*=\mathcal{E}_\mathcal{P}$), $\{r=a,\lambda = s, g(\lambda)=\lambda^{-1}f(\lambda^{-1})$ for oblate ellipsoids ($\mathcal{E}_*=\mathcal{E}_\mathcal{O}$).\footnote{Our results slightly differ from \citet{Schmalzing+1999}. 
}

Equations~(\ref{eq:ellipsoid}-\ref{eq:ellipsoid:oblateprolate}) reduce to the familiar expressions for a sphere $\mathcal{S}$ ($a=b=c$), viz. $V_0^\mathcal{S}=4\pi a^3/3$, $V_1^\mathcal{S}=2\pi a^2/3$, and $V_2^\mathcal{S}=4a/3$.

\begin{figure}
\begin{center}
\includegraphics[height=7.5cm]{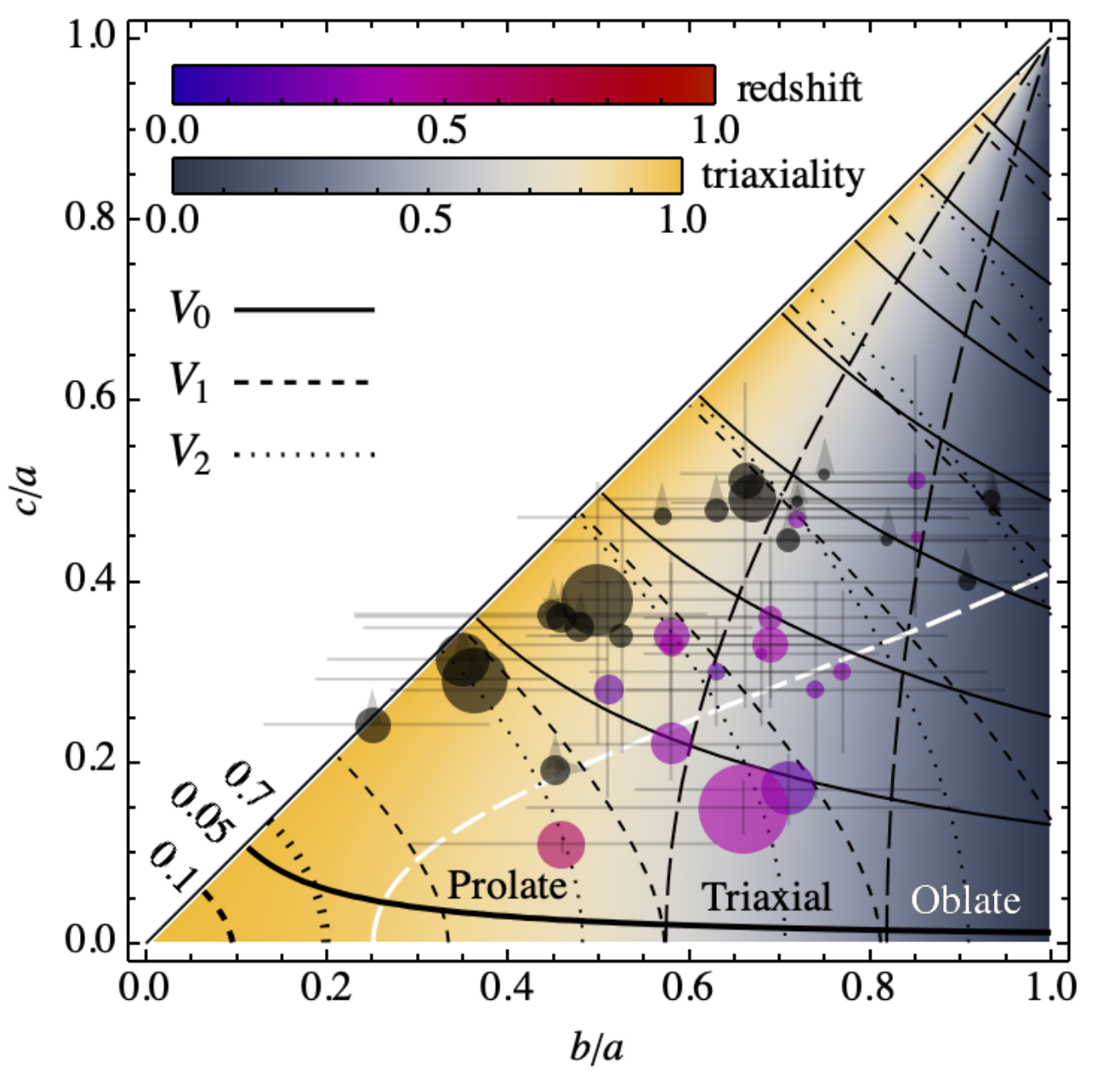}
\caption{Minkowski functionals iso-contours for an ellipsoid, $V_\mu^\mathcal{E}$, as function of the intermediate-to-major and minor-to-major axis ratio, $q=b/a$ and $s=c/a$. Volume (solid lines; $\mu=0$), surface (dashed; $\mu=1$), and integrated mean curvature (dotted; $\mu=2$) are shown for values ranging from (0.05,0.7,0.1) and increasing in steps $\Delta V_\mu=\{0.5,0.25,0.01\}~h^{\mu-3}$Mpc$^{3-\mu}$. The underlying density plot represents the triaxiality $T$. Points with error bars are CLASH clusters from \citet[colour coded by redshift; point size proportional to the mass]{Sereno+2018} and \citet[black]{Chiu+2018}, softly following the median prolatness-ellipticity relation of \citet[white long-dashed line]{DespaliGiocoliTormen2014}.
}
\label{fig:MFellipsoid}
\end{center}
\end{figure}

As illustrated in Figure~\ref{fig:MFellipsoid}, the surface area $V_1$ and the integrated mean curvature  $V_2$ are nearly degenerate with the volume $V_0$ for nearly prolate shapes or orthogonal for nearly oblate structures. They follow the trend of the triaxiality parameter $T=(1-q^2)/(1-s^2)$, which distinguishes oblate ($T\gtrsim0$) from prolate ($T\lesssim1$) structures and can be used to define three broad morphological classes \citep[][long dashed lines]{Chua+2018}. Coloured points (size proportional to the mass, colour-coded by redshift) have been obtained for the Cluster Lensing and Supernova Survey with Hubble (CLASH) clusters \citep{Sereno+2018}. Their 3D shape are constrained with a multi-wavelength analysis combining the surface mass density as determined by gravitational lensing, which probes the size in the plane of the sky, and X-ray and SZ data, to infer the radial extent \citep{Sereno2007}. With convenient priors, some less strong constraints on the 3D shape can be still determined based on lensing alone \citep[][black points]{Chiu+2018}. These points softly follows the median prolatness-ellipticity relation of \citet[white long-dashed line]{DespaliGiocoliTormen2014} that fits $\Lambda$CDM $N$-body simulations.

\subsection{Fused-balls model: merging clusters}
\label{subsec:MF:mergingballs}

Clusters are usually neither relaxed nor isolated. The complex morphology of merging clusters, or a central cluster with satellite haloes can be conveniently pictured as a group of partially overlapping balls. In this subsection, we consider first the case of a major merger ($N=2$ balls) and then the case of satellite halos ($N>2$ balls).

\paragraph*{Major mergers.}

The Minkowski functionals of two merged balls
\footnote{We denote $B_i\equiv B[\boldsymbol{x}_i,R]$ the $i$-th ball centred in $\boldsymbol{x}_i$ with radius $R$, omitted for clarity.}
$\mathcal{M}=B_1\cup B_2$ cannot be calculated like for the ellipsoid because the surface is not regular enough to uniquely define the fundamental form. Instead, they can be calculated using additivity, $V_\mu(B_1\cup B_2)=V_\mu(B_1)+V_\mu(B_2)-V_\mu(B_1\cap B_2)$. For two merged balls with unequal radii $R$ and $r\leqslant R$ and centres at distance $d\leqslant R+r$, the volume and surface area are trivial \citep[see also][]{GibsonScheraga1987}, while the integrated mean curvature can be calculated using the Steiner formula; see Appendix~\ref{sec:appendix:mergingspheres}.
One finally obtains
\begin{subequations}\label{eq:mergingspheres}
\begin{align}
V_0^\mathcal{M}&= \frac{{2}\pi}{3}\left(R^3+r^3-\frac{1}{8}d^3\right)
	 + \frac{\pi}{2}(R^2+r^2)d + \frac{\pi}{4d}(R^2-r^2)^2, \label{eq:mergingspheresV0} \\
V_1^\mathcal{M}&= \frac{\pi}{3}(R^2+r^2)
	 + \frac{\pi}{6}(R+r)d + \frac{\pi}{6d}(R-r)(R^2-r^2), \label{eq:mergingspheresV1} \\
V_2^\mathcal{M}&= \frac{2}{3}(R+r+d)
	-\frac{\psi}{3}d\sqrt{2\frac{R^2+r^2}{d^2}-1-\left(\frac{R^2-r^2}{d^2}\right)^2},
	\label{eq:mergingspheresV2}
\end{align}
\end{subequations}
with $\cos\psi=(R^2+r^2-d^2)/2Rr$. These equations are defined for non-trivial merging, i.e. as long as the two spheres overlap with no total embedding ($B_1\cap B_2\neq\emptyset$ and $B_2\nsubseteq B_1$, i.e. $R-r\leqslant d$); for non-overlapping spheres the correct expression is recovered setting $d=R+r$.

\begin{figure}
\begin{center}
\includegraphics[height=7.1cm]{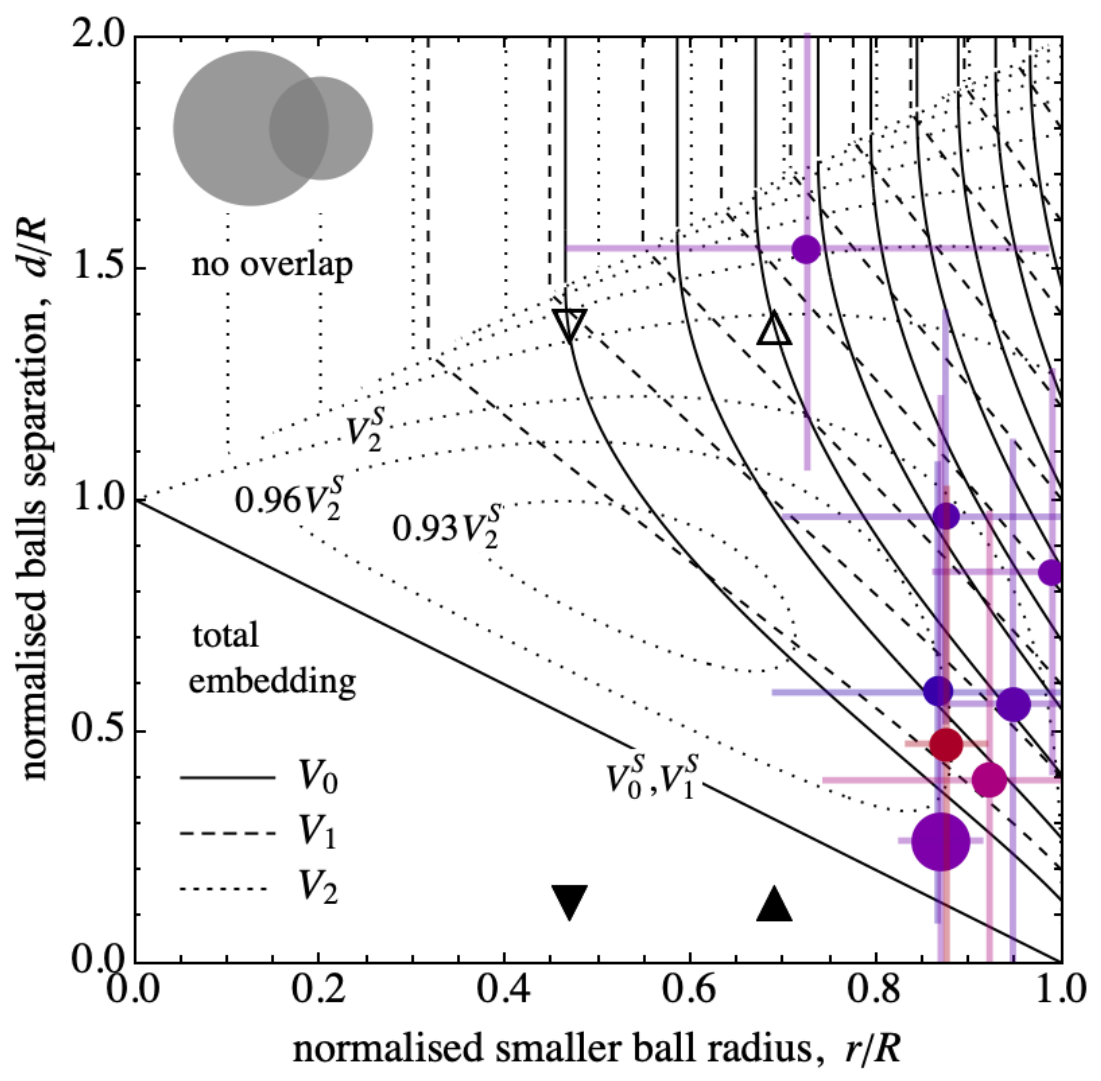}
\caption{Minkowski functionals iso-contours for major mergers, $V_\mu^\mathcal{M}$ (Eqs.~\ref{eq:mergingspheres}), as function of the smaller ball radius $r$ and distance $d$ from the major ball with radius $R$. Volume (solid lines), surface (dashed), and integrated mean curvature (dotted) levels increase by 10\% moving top-right from the lower values attained for a single ball, $V_\mu^\mathcal{S}$ (for $V_2^\mathcal{M}$, also the contours corresponding to 93 and 96\,\% of $V_2^\mathcal{S}$ are shown). The lower (`total embedding') and upper (`no overlap') triangular regions account for trivial morphologies of one and two isolated balls, respectively. Points indicate major mergers from the LC$^2$ cluster catalogue \citep[][colour-coded by redshift as in figure~\ref{fig:MFellipsoid}, size proportional to $R_{200\mathrm{c}}$]{Sereno2015}. Filled (empty) symbols designate a merging subclump at the pericenter (splashback) for a system at $z=0.3$; see Table~\ref{tab:LC2-phases-twoballs}.
}
 \label{fig:MFmergingballs}
\end{center}
\end{figure}

The results are illustrated in Figure~\ref{fig:MFmergingballs} as function of the radius of the smaller ball and separation between the centres, both normalised to the radius of the larger ball.
Note that for major ($r\lesssim R$) and advanced ($d\ll R$) mergers, $V_0$ and $V_1$ are nearly degenerate. As reference, the values for the major-mergers ($r\sim R$) from the LC$^2$ catalogue \citep{Sereno2015} calculated assuming a flat $\Lambda$CDM cosmology with $\Omega_\mathrm{m}=0.3$ and $h=0.7$ and $R\equiv R_{200\mathrm{c}}$ are shown, along with the characteristic splashback \citep{Diemer+2017} and pericenter values estimated for binary systems at redshift $z=0.3$ with main halo mass $M_{200\mathrm{c}}=10^{14}h^{-1}\mathrm{M}_{\sun}$ and secondary halo with 3 or 10 times smaller mass;\footnote{$M_{200\mathrm{c}}$ denotes the mass enclosed within a sphere of radius $R_{200\mathrm{c}}$ with mean over-density 200 times the critical density.} 
 see Table~\ref{tab:LC2-phases-twoballs}.

For balls with equal radius ($R=r$) the Minkowski functionals of the resulting body $\mathcal{M}_\mathcal{P}$ are well-known,
\begin{subequations}
\begin{align}
V_0^{\mathcal{M}_\mathcal{P}} &= \frac{4\pi}{3}R^3\left(1 + \frac{3d}{4R} - \frac{1}{16}\frac{d^3}{R^3}\right),
 \\
V_1^{\mathcal{M}_\mathcal{P}} &= \frac{2\pi}{3}R^2\left(1 + \frac{d}{2R}\right),
 \\
V_2^{\mathcal{M}_\mathcal{P}} &= \frac{4R}{3}\left(1+\frac{d}{2R}\right)-\frac{2R}{3}\sqrt{1-\frac{d^2}{4R^2}}\arccos\left(1-\frac{d^2}{2R^2}\right).
\end{align}
\end{subequations}
with limits $V_0^\mathcal{S}=4\pi R^3/3$, $V_1^\mathcal{S}=2\pi R^2/3$, and $V_2^\mathcal{S}=4R/3$ when $d=0$, i.e. when $\mathcal{M}_\mathcal{P}$ becomes a sphere, $\mathcal{S}$.

\begin{table}
\centering
\caption{LC$^2$ merging clusters \citep{Sereno2015}: parameters of two-balls model ($R\equiv R_{200\mathrm{c}}$, flat $\Lambda$CDM cosmology). For comparison (lower part of the table), indicative values for the splashback and pericenter phase of some major merger.
Lengths in $h^{-1}$Mpc.
}
\label{tab:LC2-phases-twoballs}
\resizebox{1.01\hsize}{!}{
	\begin{tabular}{lcccc}
	\hline
	Name & redshift & $R$ & $ r$ & $d$ \\
	\hline
	Abell 1750 & 0.0678 & $0.98\pm0.19$ & $0.85\pm0.20$ & $0.57\pm0.06$ \\
	Abell 901 & 0.16   & $0.88\pm0.15$ & $0.77\pm0.19$ & $0.85\pm0.08$ \\
	Abell 115 & 0.197  & $1.13\pm0.10$ & $1.07\pm0.11$ & $0.63\pm0.06$ \\
	Zw Cl2341 & 0.27   & $0.87\pm0.15$ & $0.86\pm0.15$ & $0.73\pm0.07$ \\
	Abell 1758 & 0.28   & $0.95\pm0.26$ & $0.68\pm0.17$ & $1.46\pm0.14$ \\
	Bullet Cluster & 0.296  & $1.91\pm0.09$ & $1.66\pm0.09$ & $0.50\pm0.05$ \\
	MACS J0025 & 0.5842 & $1.15\pm0.28$ & $1.06\pm0.23$ & $0.45\pm0.05$ \\
	CLJ0102-4915 & 0.87   & $1.10\pm0.05$ & $0.96\pm0.05$ & $0.52\pm0.05$ \\
	\hline
	\multicolumn{5}{c}{$M_1=10^{14}h^{-1}\mathrm{M}_{\sun}$, $M_2=M_1/10$:} \\
	splashback $\triangledown$ & 0.3 & 1.47 & 0.68 & 2.03 \\
	pericenter $\blacktriangledown$  & 0.3 & 1.47 & 0.68 & 0.20 \\
	\multicolumn{5}{c}{$M_1=10^{14}h^{-1}\mathrm{M}_{\sun}$, $M_2=M_1/3$:} \\
	splashback $\vartriangle$ & 0.3 & 1.47 & 0.68 & 2.03 \\
	pericenter $\blacktriangle$ & 0.3 & 1.47 & 0.68 & 0.20 \\
	\hline
	\end{tabular}
}
\end{table}

\paragraph*{Multiple mergers.}

\begin{figure*}
\begin{center}
\includegraphics[width=1.0\textwidth]{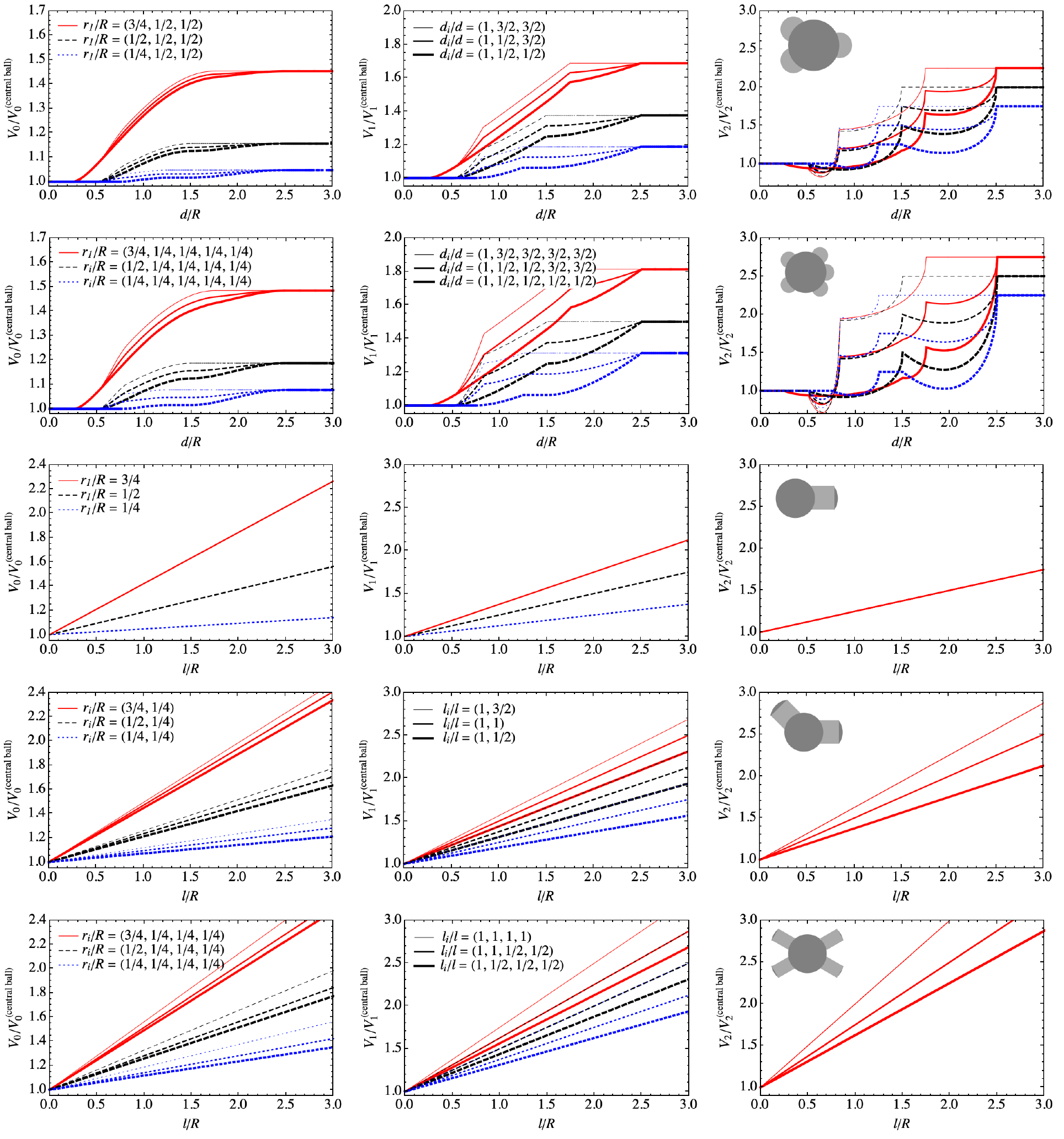}
\caption{Minkowski functionals of merging models $\mathcal{M}_n$ with $n=3,5$ satellites (rows 1-2 from the top) and spiky models $\mathcal{S}_n$ with $n=1,2,4$ filaments (rows 3-4-5), illustrated by topologically equivalent bodies in the top left corner of the right panels. Left-to-right: volume, surface, and integrated curvature, normalised to the values of the central ball. In $\mathcal{M}_n$ the satellite balls $B_i$ have different radius $r_i$ and are at distance $d_i\propto d$ from the central ball $B$ (see legends); lines with increasing thickness would represent subsequent stage of merging, with satellites going closer to $B$. In $\mathcal{S}_s$ the cylindric filaments have bases of radius $r_i$ and length $\ell_i\propto \ell$ (see legends); thicker lines represent later stages of gravitational evolution. For the $\mathcal{S}_s$ models $V_2$ does not depend on the radius of filaments but only on their length. Lengths are in units of the central ball radius $R$.}
\label{fig:MFmergers+spiky}
\end{center}
\end{figure*}

Equations~(\ref{eq:mergingspheres}) can be extended to the easiest configuration for multiple merging, i.e. a central ball (halo) $B$ of radius $R$ intersecting $n$ smaller balls (satellites) $B_i$ of radius $r_i$, with centres at distance $d_i$ from the centre of $B$ and not mutually intersecting ($B_i\cap B_j = \emptyset$; $i,j=1,\dots,n$; $N=n+1$). The Minkowski functionals of the simply-connected resulting body $\mathcal{M}_n = B \cup \bigcup_{i=1}^n B_i$ (so $\mathcal{M}_1\equiv\mathcal{M}$) are trivially obtained by additivity; see Appendix~\ref{sec:appendix:MultiMergersModel}. The two top-line panels in Figure~\ref{fig:MFmergers+spiky} illustrate the results for $n=3$ and 5.

\subsection{Spiky model: filaments feeding clusters}
\label{subsec:MF:clusters+filaments}

Massive halos form at the highest density nodes of the cosmic web. Even in absence of major mergers, dark matter is continuously accreting along filaments connecting the nodes \citep[e.g.][]{Eckert+2015,Connor+2018}. We can approximate such spiky geometry by a ball $B$ of radius $R$ attached to $n$ distinct i.e. not mutually intersecting cylinders $C_i$ $(i=1,\dots,n)$ radially joined to $B$, each with length $\ell_i$ and basis with radius $r_i$ lying on the surface of $B$. Using additivity, the Steiner formula, and Equation~(\ref{eq:volwedgetorus}) the Minkowski functionals of the resulting body $\mathcal{\bar{S}}_n=B\cup\bigcup_{i=1}^n C_i$ are
\begin{subequations}\label{eq:stellarcylinders}
\begin{align}
~~V_0^{\mathcal{\bar{S}}_n}&= \frac{\pi}{3}(2-n)R^3 + \pi \sum_i \left[r_i^2 \ell_i +\frac{p_i}{3}(3R^2-2Rp_i-p_i^2)\right], \\
V_1^{\mathcal{\bar{S}}_n}&= \frac{\pi}{3}(2-n)R^2 + \frac{\pi}{6} \sum_i (r_i^2+2r_i\ell_i-2Rp_i), \\
V_2^{\mathcal{\bar{S}}_n}&= \frac{2}{3}(2-n)R + \frac{1}{3} \sum_i \left( \ell_i + 2p_i + r_i\arcsin\frac{r_i}{R}\right),
\end{align}
\end{subequations}
in which $p_i=(R^2-r_i^2)^{1/2}$ is the distance from the centre of $B$ to the $i$-th spherical cap bounded by the cylinder $C_i$. The condition $C_i\cap C_j = \emptyset$ is possible if approximately $\sum_i r_i^2\lesssim4 R^2$.

Equations~(\ref{eq:stellarcylinders}) are simpler but still keep the essential information if the free heads of cylinders are not flat but spherical caps with the same curvature radius as the central ball, i.e. $\mathcal{L}_i = B\cap C_i$, so that $V_\mu(\mathcal{S}_n)=V_\mu(B)+\sum_iV_\mu(C_i)$. The Minkowski functionals are then
\begin{subequations}\label{eq:stellarcylinders2}
\begin{align}
~~V_0^{\mathcal{S}_n} &= \frac{4\pi}{3}R^3 + \pi \sum_i r_i^2 \ell_i\,, \\
V_1^{\mathcal{S}_n} &= \frac{2\pi}{3}R^2 + \frac{\pi}{3} \sum_i r_i\ell_i\,, \\
V_2^{\mathcal{S}_n} &= \frac{2}{3}R + \frac{1}{3} \sum_i \ell_i\,.
\end{align}
\end{subequations}
The bottom rows of Figure~\ref{fig:MFmergers+spiky} illustrate results for $n=1,2,4$.

\begin{figure}
\begin{center}
\includegraphics[width=7.7cm]{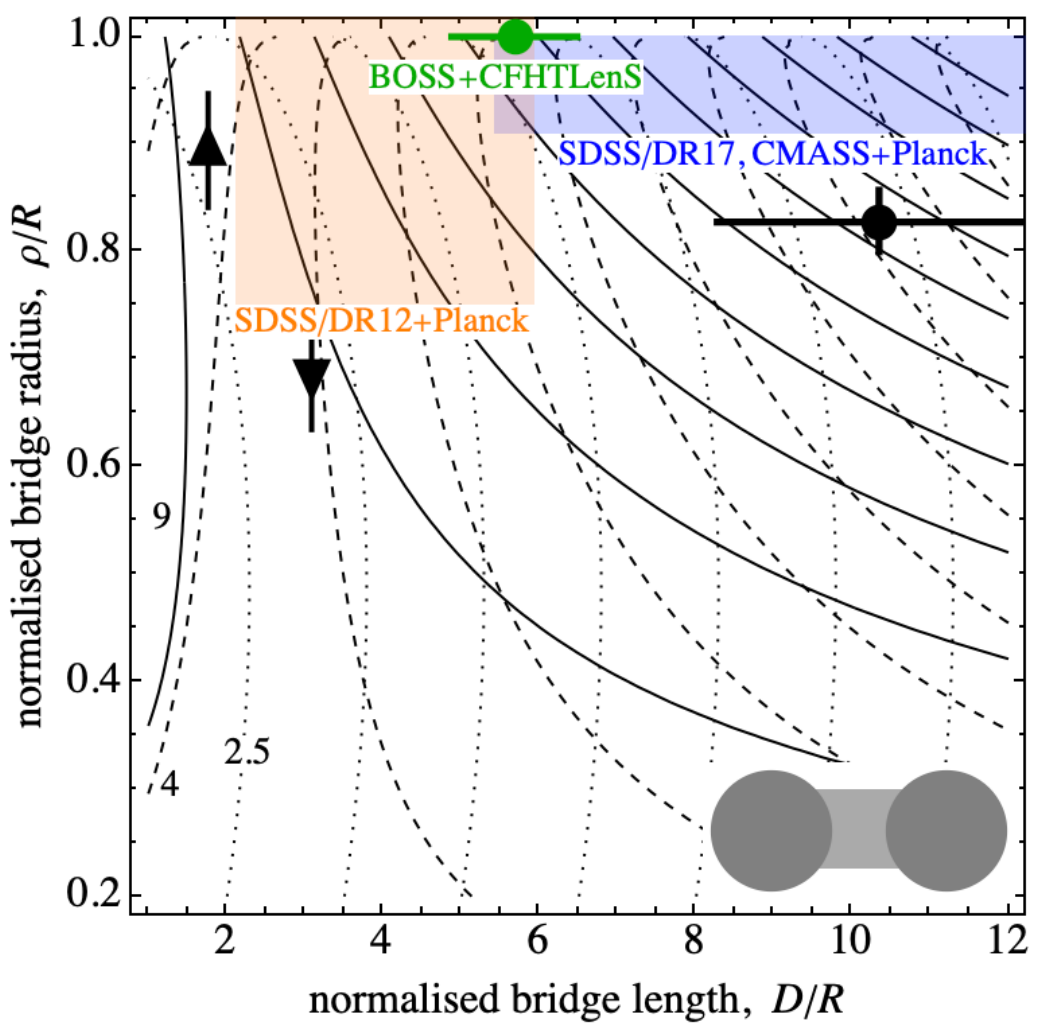}
\caption{Minkowski functionals iso-contours for the dumbbell model $\mathcal{D}$ with balls of same radius $R$ as function of distance $D$ between the balls' centres and of radius $\rho$ of the cylindric bridge. Volume (solid lines; $\mu=0$), surface (dashed; $\mu=1$), and integrated mean curvature (dotted; $\mu=2$) are shown for values ranging from the smaller as indicated and in steps $\Delta V_\mu=\{2,0.5,0.5\}$~Mpc$^{3-\mu}$ moving rightward. Data points are described in Table~\ref{tab:bridges}.}
 \label{fig:MFdumbbell}
\end{center}
\end{figure}


\subsection{Dumbbell model: cluster-pair bridge}\label{subsec:MF:dumbbell}

\begin{table}
\caption{Observed cluster-pair bridges of single (upper table) or stacked (lower table) systems. The compilation is restricted to clusters pairs with similar radius $R\approx r=r_{200}$, separated by $D$ and with filament radius $\rho$. Lengths in $h^{-1}$Mpc.}
\label{tab:bridges}
\centering
\resizebox{\hsize}{!}{\begin{tabular}{lcccc}
\hline
Cluster/data set & redshift & $R$ & $D$ & $\rho$ \\
\hline
A222 - A223 ($\bullet$) & 0.21 & 1.2 & $15\pm3$ & 0.6 \\
A399 - A401 ($\blacktriangle$) & 0.073 & $1.70$ & 3 & $1.52\pm0.09$ \\
A21 - PSZ2 G114.9 ($\blacktriangledown$) & 0.094 & $1.36$ & 4.2 & 0.92 \\
\hline
SDSS/DR17 (LRG) & 0.2--0.5 & 0.5--1 & 6--14 & $\gtrsim1$ \\
BOSS + CFHTLenS & 0.3--0.6 & 1.25 & $7.1\pm1$ & 1.25 \\
SDSS/DR12 + Planck & 0--0.4 & 1.35 & 6--10 & $\leqslant 0.5$ \\
CMASS + Planck & $\sim0.55$ &0.5--1 & 6--14 & $\leqslant 2.5$ \\
\hline
\end{tabular}
}
\end{table}

Cluster of galaxies may reside in superclusters still not in equilibrium. In the simpler configuration, major haloes are connected through thick filaments \citep[e.g.][Table~\ref{tab:bridges}, rows 1-3]{Werner+2008,Dietrich+2012,A399A401-Planck,A399A401-Planck2}, also detected by stacking techniques \citep[][Table~\ref{tab:bridges}, rows 4-7]{Clampitt+2016,EppsHudson2017,Tanimura+2019,deGraaff+2019}.

\begin{figure*}
\begin{center}
\includegraphics[width=0.9\textwidth]{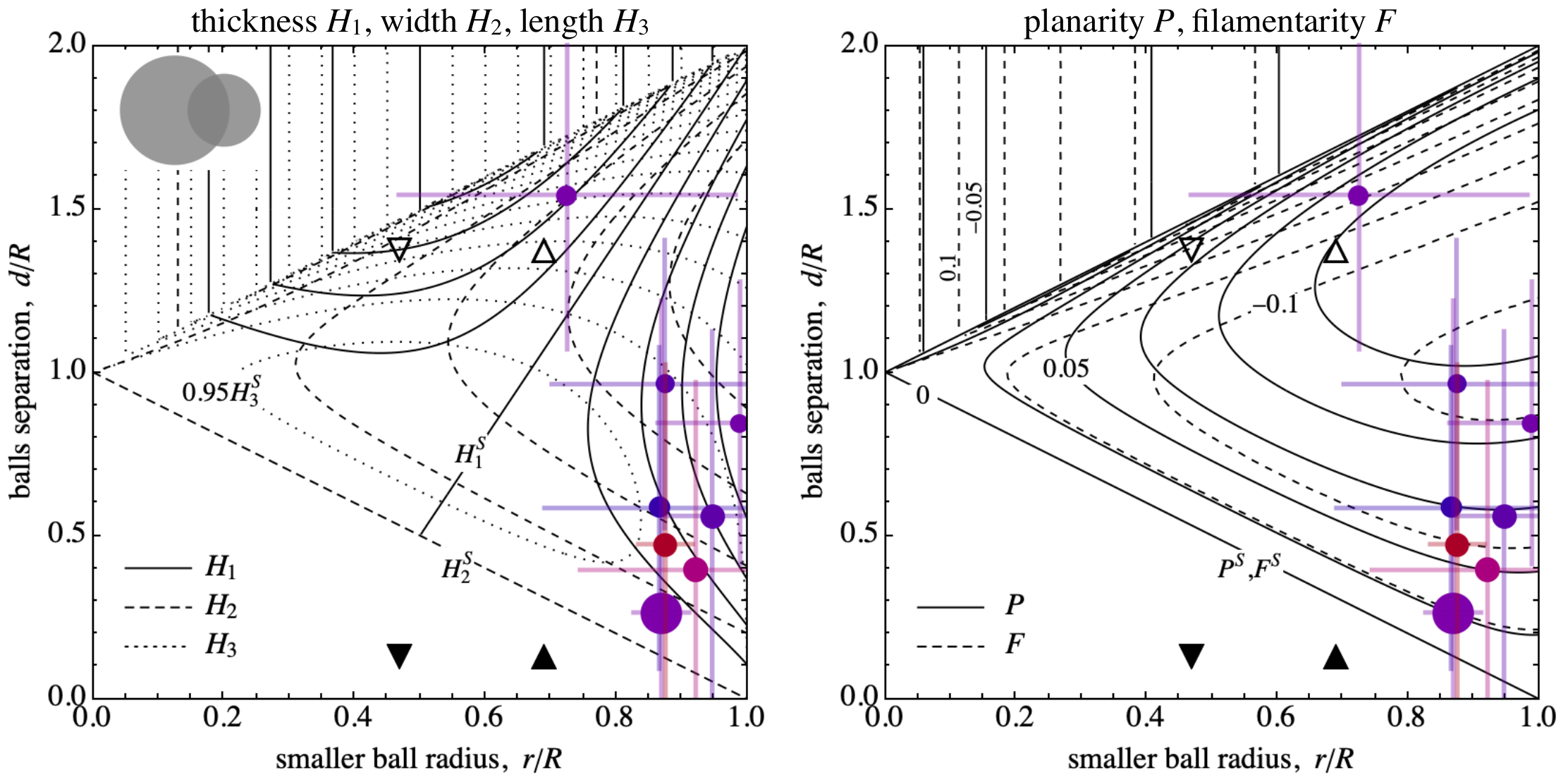}
\caption{Shapefinders isocontours for merger model $\mathcal{M}$, where two balls with radius $R$ and $r$ are separated by $d$.
\textit{Left:} $H_1,H_2,H_3$ isocontours in units of the values attained for a unit ball $\mathcal{S}$, i.e. $H_1^\mathcal{S}=H_2^\mathcal{S}=H_3^\mathcal{S}=1$, increasing by $\Delta H_i/H_i^\mathcal{S}=(0.025,0.1,0.05)$ rightward. $H_1$ and $H_2$ are minimum when the two balls do not overlap, respectively for $r/R\sim0.89$ and 0.41; $H_3$ is minimum for non-trivial overlap, at $(r/R,d/R)\approx(0.54,0.84)$.
\textit{Right:} planarity (filamentarity) isocontours range in $[-0.025,0.125]$ ($[-0.15,0.3]$) in steps of 0.025 (0.05), crossing the vanishing value valid for a ball or through total embedding. Symbols as in Figure~\ref{fig:MFmergingballs}.
}
\label{fig:H123PFmergingballs}
\end{center}
\end{figure*}

The morphology of an axially-symmetric body defined by two balls connected by a cylinder, $\mathcal{D}=B_1\cup B_2\cup C$, can be deduced from the previous equations using additivity and noting that the sum of the Minkowski functionals of the two spherical caps chopped by the cylinder bases, $\mathcal{L}_{1,2}=B_{1,2}\cap C$, are equivalent to the Minkowski functionals of the lens $\mathcal{L}=\mathcal{L}_1\cup \mathcal{L}_2$ as reported in Appendix~\ref{sec:appendix:mergingspheres}. After some algebra and recognising the two-fused balls model, one obtain $V_\mu(\mathcal{D})=V_\mu^\mathcal{M}+V_\mu(C)$. The exact though cumbersome mathematical expression combines Equations~(\ref{eq:mergingspheres}) for two balls of radius $R$ and $r$ separated by an effective distance $d=(R^2-\rho^2)^{1/2}+(r^2-\rho^2)^{1/2}$, and the well-known Minkowski functionals of a cylinder with circular basis of radius $\rho$ and height $D-d$, with $D$ the actual distance between the centres of $B_1$ and $B_2$. An illustrative example of Minkowski functionals iso-contours for balls with same radius $R$ is shown in Figure~\ref{fig:MFdumbbell} as function of length and radius of the cylindric bridging filament. For relatively small bridge lengths ($D \sim 2 R$), the functionals are quite degenerate. For larger radii, degeneracy is broken. A compilation of systems that can be approximated by this geometry are reported for comparison; see Table~\ref{tab:bridges}.

A more advanced configuration is obtained by replacing the cylinder by a truncated cone $P$ with circular bases of radius $\rho_1$ and $\rho_2$ and height $h=D-(R^2-\rho_1^2)^{1/2}-(r^2-\rho_2^2)^{1/2}$; see Appendix~\ref{sec:appendix:dumbbell}. The Minkowski functionals are $V_\mu(\mathcal{D}_P)=V_\mu(B_1)+V_\mu(B_2)+V_\mu(P)-V_\mu(\mathcal{L}_1)-V_\mu(\mathcal{L}_2)$, with the functionals for $P$ obtained from Equations~(\ref{eq:dumbbell1}-\ref{eq:dumbbell3}). It is not difficult to further generalise this model by adding two additional cylindric filaments that protrude from the two haloes in opposite directions. These two haloes can be regarded as local clumps of matter embedded in a single, bent cosmic filament similar to the A3016-A3017 system \citep{A3016A3017-Chon+2019}. Finally, note that a pile of truncated cones with matching bases can describe axially-symmetric filaments with varying thickness, well-suited for systems such as the one recently reported by \citet{Umehata+2019} and \citet{HerenzHayesScarlata2020}.

\begin{figure*}
\begin{center}
\includegraphics[width=0.95\textwidth]{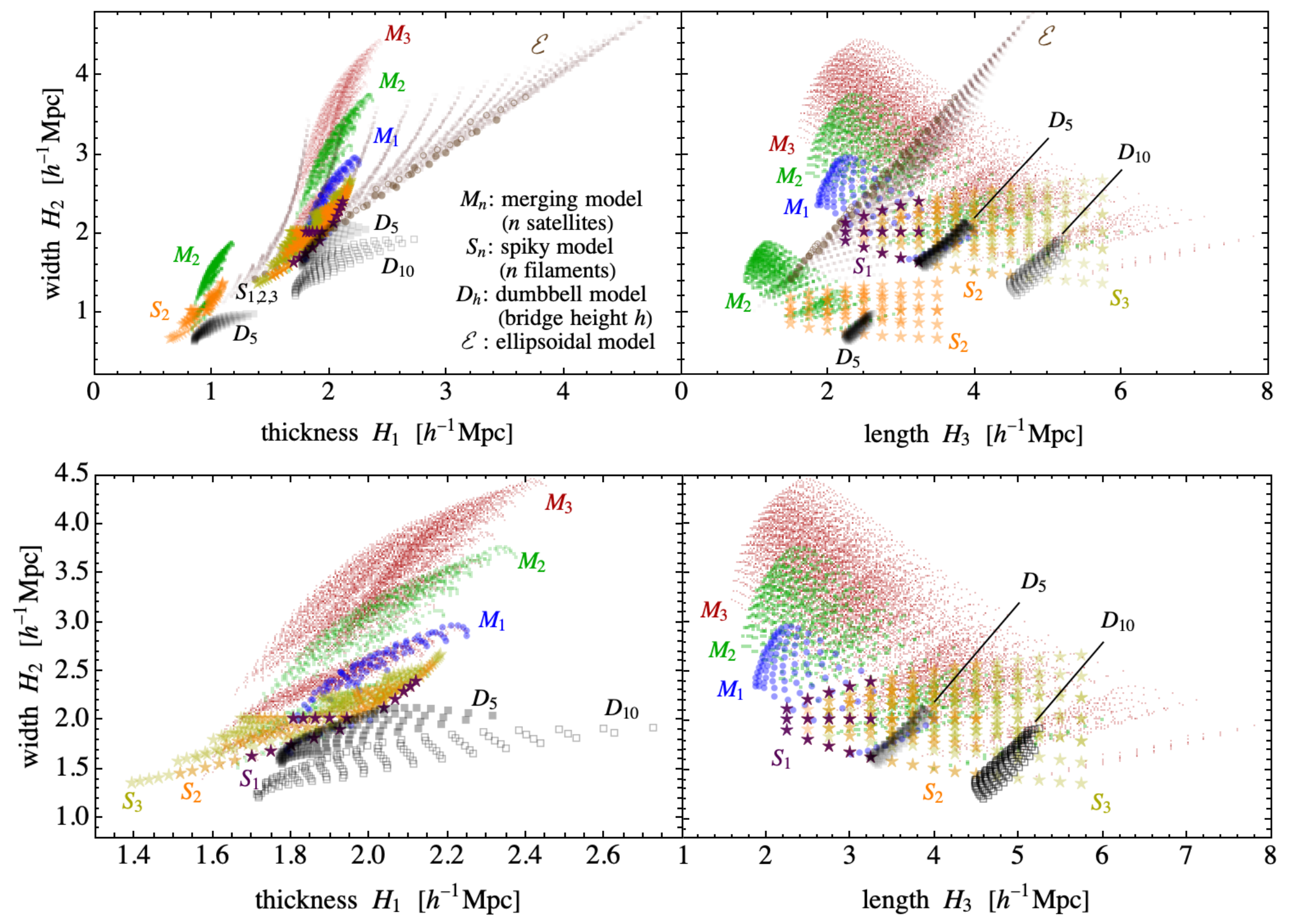}
\caption{Classification by shapefinders $(H_1,H_2,H_3)$ of ellipsoidal triaxial, merging, spiky, and dumbbell models in two projections (left and right panels; see \S~\ref{sec:classification} for values).
\textit{Top:} merging, spiky, and dumbbell models have central or major ball with radius
$R=1$ or $2h^{-1}$Mpc. Correspondingly, they have $H_1\approx1$ or $2h^{-1}$Mpc and $H_2\gtrsim1$ or $2h^{-1}$Mpc.
\textit{Bottom:} zoom on models with major ball with $R=2h^{-1}$Mpc.
}
\label{fig:H123}
\end{center}
\end{figure*}

\vspace{-0.3cm}
\subsection{Mass assembly history and morphology} 
During the late stage of evolution before virialisation, satellite haloes are closer to the main halo. This tends to accrete mass at merging rate and with time-scale depending on the epoch, initial mass and statistics of the primordial density field \citep{Bond+1991}, mass and kinematic of sub-haloes \citep[e.g.][]{Zhao+2003}, and tidal forces 
 \citep{LapiCavaliere2011}, which are possibly conditioned by dark-energy \citep{Pace+2019}. The filamentary structures feeding clusters tend to become shorter and thinner \citep[e.g.][]{Cautun+2014}, and the connectivity of the more massive hence largest and latest formed haloes decreases over time \citep{Choi+2010,CodisPogosyanPichon2018,Kraljic+2020}.
Since Minkowski functionals account for the non-trivial geometrical and topological content of fused bodies despite their evolutionary stage, relaxed or not, one expects that they correlate with the dynamical state of galaxy clusters. This claim is supported by the results we obtained with idealised models.

As shown in Figure~\ref{fig:MFmergers+spiky} (top panels), while the volume of merging models $\mathcal{M}_n$ is mainly sensitive to the relative size (radius) of the satellites, area and integrated mean curvature strongly depend also on their relative distance from the main halo, lifting the degeneracies. Overall, late-time structures are more compact i.e. occupy smaller volume, cover smaller area, and have smaller intrinsic curvature than at early time (later stages of the gravitational evolution are represented by thicker lines and by solid-dashed-dotted sequence).

The morphology captured by Minkowski functionals for spiky models $\mathcal{S}_n$ (Figure~\ref{fig:MFmergers+spiky}, bottom rows) is similar to merging models: regardless the number of filaments attached to the central ball, the volume primarily depends on the thickness (radius) of filaments, the area is likewise sensitive to the relative length of filaments, $\ell_i$, while integrated mean curvature only depends on $\ell_i$. Again, the overall amplitude of Minkowski functionals decreases for late-time morphologies, converging towards the values of the central main cluster.

A numerical study based on $N$-body simulations is needed to quantitatively assess the correlation between the full set of Minkowski functionals and the relaxation state of these structures. It will be of interest to evaluate the ability of these statistics to distinguish between `stalled' and `accreting' haloes, which are located at the nodes of a network of respectively thin and thick filaments feeding them \citep{BorzyszkowskiPorciani+2017}.

\vspace{-12pt}
\section{Classification by shapefinders}\label{sec:classification}

\citet{SahniSathyaprakashShandarin1998} introduced the thickness $H_1=V_0/2V_1$, width $H_2=2V_1/\pi V_2$, and length $H_3=3V_2/4$ of isodensity contours, dubbed shapefinders, to investigate in a non-parametric way the size and shape of the matter density field above or below a given threshold on large scales.
\citet{ShandarinShethSahni2004} used these statistics for voids and superclusters.
We employ the shapefinders to attempt a classification of the morphology of galaxy clusters.
The shapefinders are geometrically and physically motivated and can characterise all the different phases of the merging accretion history, or, in a complementary view, all the halo configurations that populate the universe at a single cosmic time.

The dependence of shapefinders on the parameters of models is illustrated only for the two-fused ball model, $\mathcal{M}$, as prototype for major mergers and fully accounted for by two parameters only, i.e. the ratio of balls radii $r/R$ and the distance between balls in units of the major ball radius, $d/R$. Figure~\ref{fig:H123PFmergingballs} (left panel) suggests that major-merging clusters from the LC$^2$ catalogue, which share similar geometric scales $r$ or $d$, can be distinguished by the values of $H_1$, $H_2$, and $H_3$, whose iso-contours are markedly orthogonal in different part of the $(r,d)$ parameter space. For reference, the two major-merging systems with secondary haloes orbiting at the splashback radius of the main halo and having 3 and 10 times smaller mass (upper and lower empty triangles, respectively) differ by about 16, 12, and 6 percent in volume, surface, integrated mean curvature, corresponding to a 4, 16, and 6 percent difference in $H_1$, $H_2$, and $H_3$, respectively.

The so-called planarity $P=(H_2-H_1)/(H_2+H_1)$ and filamentarity $F=(H_3-H_2)/(H_3+H_2)$ are less suitable shapefinders to classify the systems considered in this study, especially the `stellar' models $\mathcal{M}_n$ and $\mathcal{S}_n$; here the words `planarity' and `filamentarity' are equivocal. Nonetheless, for the two-fused ball model $P$ and $F$ can differ by about $\pm0.15$ from zero, which is the value for a ball. A Blaschke diagram based on $(P,F)$ \citep[see e.g.][]{Schmalzing+1999} could be therefore an alternative interesting diagnostic to classify more realistic systems.

\begin{figure*}
\begin{center}
\includegraphics[width=0.98\textwidth]{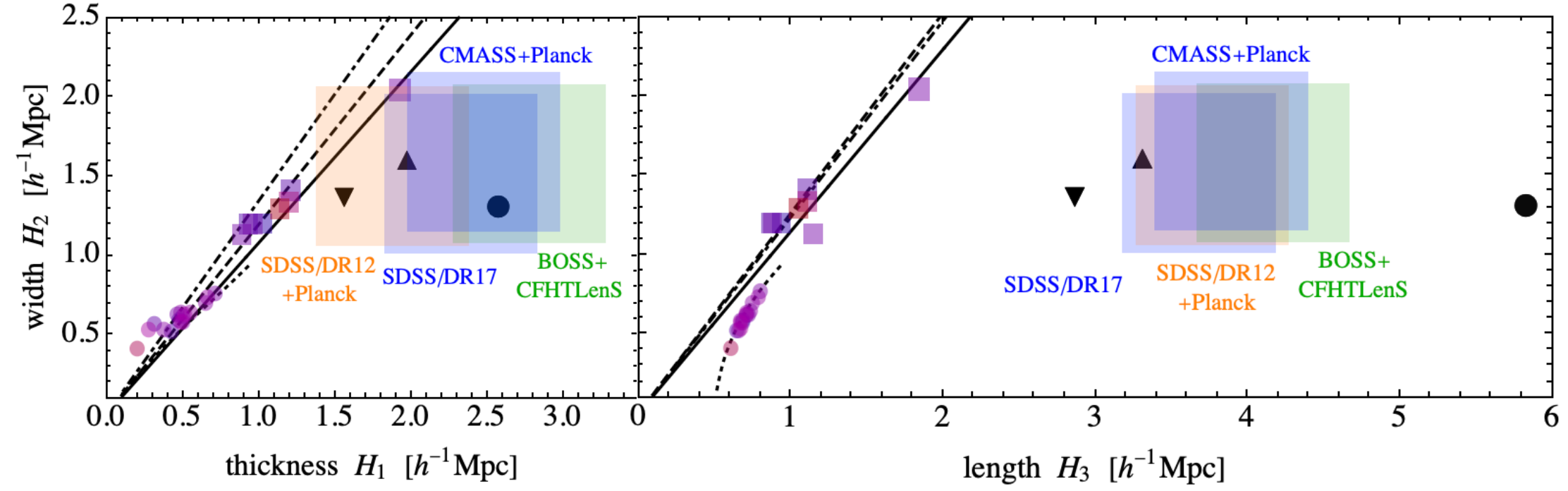}
\caption{Example of shapefinders classification of observed systems (projections as in Figure~\ref{fig:H123}): triaxial haloes \citep[][disks, colour-coded by  redshift as in Figure~\ref{fig:MFellipsoid}]{Sereno+2018}, major mergers \citep[LC$^2$ catalogue by][squares, colour-coded by redshift as in Figure~\ref{fig:MFmergingballs}]{Sereno2015}, dumbbell systems (black symbols and shaded regions, see Table~\ref{tab:bridges} and Figure~\ref{fig:MFdumbbell}).
For reference, major merging models $\mathcal{M}$ are shown, all having main halo with radius $R=0.1-3h^{-1}$Mpc and secondary halo with radius $r$ located at distance $d$ such that $(r,d)=(R, 0.3R)$ (i.e. close haloes with same radius; solid line), $(0.5 R, R)$ (dashed), and $(R, 1.5R)$ (i.e. distant haloes with same radius; dot-dashed). Triaxial haloes by \citet{Sereno+2018} nicely fit the ellipsoidal prolate model with $(a,b,c)=(1~h^{-1}\mathrm{Mpc},b,b)$, $b\in[0.1,1]~h^{-1}$Mpc in the $(H_2,H_3)$ plane (dotted line) but not in the $(H_1,H_2)$ plane.
}
\label{fig:H123data}
\end{center}
\end{figure*}

The geometrical models presented in Section~\ref{sec:MF} illustrate the potentiality of the classification scheme based on $(H_1,H_2,H_3)$, which can be applied to clusters of galaxies or any astrophysical or physical system with non-trivial geometry. Figure~\ref{fig:H123} shows the three projections of the $(H_1,H_2,H_3)$ parameter space populated with triaxial ellipsoids $\mathcal{E}$ (with axes $a,b,c\in[1, 5]~h^{-1}$Mpc), merging models $\mathcal{M}_n$ with $n=1,2,3$ satellites (balls with radius $r_i\in[R/2,R]$ at distance $d\in[R,2R]$ from the central ball with radius $R=1,2h^{-1}$Mpc), spiky models $\mathcal{S}_n$ with $n=1,2,3$ filaments (cylinders with radius $r_i\in[R/4,3R/4]$ and length $\ell_i\in[1,5]h^{-1}$Mpc, feeding a central ball with radius $R=1,2~\,h^{-1}$Mpc), and dumbbell models $\mathcal{D}_h$ (major ball with radius $R=1,2~h^{-1}$Mpc, minor ball with radius $r\in[R/2,R]$, cylindric bridge with radius $\rho\in[R/4,3R/4]$ and height $h=5,10~h^{-1}$Mpc).\footnote{Length units are here irrelevant since only ratios do matter; in Figure~\ref{fig:H123} we adopt $h^{-1}$Mpc as common practice in cosmology.}

The triaxial ellipsoids $\mathcal{E}$ (shown only in the left panels) extend over the broadest region of the parameter space, quite well-separated from the other models. For fixed width $H_2$, the maximum thickness $H_1$ and minimum length $H_3$ is achieved for prolate and oblate ellipsoids, which are almost superposed. Instead, as shown in Figure~\ref{fig:MFellipsoid}, the same value of the Minkowski functionals corresponds to different values of the triaxiality parameter, viz. $V_\mu$ are orthogonal to $T$ so bringing less discriminating power than the shapefinders.

All but the $\mathcal{E}$ models are approximately centred around a value of $H_1$ that is equal to the radius $R$ of the central or major ball. Disregarding the unavoidable degeneracies between models, for fixed value of $R$ (see right panels, showing only models with $R=2h^{-1}$Mpc) there is a clear trend of $H_2$ that increases with the number $n$ of satellites in merging models $\mathcal{M}_n$, while it only mildly decreases with the number $n$ of cylindric filaments (connectivity) of spiky models $\mathcal{S}_n$. The connectivity is instead more evident in the $(H_2,H_3)$ plane, increasing on average with $H_3$. A similar trend occurs for the $\mathcal{M}_n$ models, with larger integrated mean curvature or $H_3$ occurring for systems with more satellites.

The Minkowski functionals support these conclusions. As suggested by Figure~\ref{fig:MFmergers+spiky} (rows 1-2), while the volume ($V_0$) and, to smaller extent, the surface ($V_1$) of merging systems $\mathcal{M}_n$ mainly inform about the size of the central ball and that of the largest satellite, the integrated mean curvature ($V_2$) is sensitive also to the smaller satellites even when $n$ is small, catching both their size and distance from the central ball. For $\mathcal{S}_n$ models (rows 3-5), $V_0$ and $V_1$ equally respond to the thickness and lengths of the filaments, while 
 the slope of $V_2$ as function of the typical length increase on average with the connectivity $n$; in this case the classification in the $(H_2,H_3)$ plane seems more selective.

Dumbbell models $\mathcal{D}_h$ attain the largest value of $H_1$, which increases with the length $h$ of the bridge. Consistently with $V_\mu$ (see Figure~\ref{fig:MFdumbbell}), the width $H_2$ is mainly sensitive to the radius of the smaller ball, while the length $H_3$ is strongly responsive to the bridge length almost regardless the other scales of the dumbbell.

Figure~\ref{fig:H123data} illustrates the ability of shapefinders to separate the observed systems presented in the precedent sections. Triaxial ellipsoids by \citet[][disks]{Sereno+2018}, major mergers of the LC$^2$ catalogue \citet[][squares]{Sereno2015}, and cluster-pair bridges (see Table~\ref{tab:bridges}, symbols and large squares) nicely occupy different positions in the parameter space.

\vspace{-12pt}
\section{Discussion and Conclusions}
\label{sec:conclusions}
The forthcoming generation of imaging and spectroscopic surveys carried out with DESI, WEAVE, 4MOST, Rubin Observatory, Euclid, Roman Space Telescope, eROSITA, or SKA will collect thousands of new galaxy clusters and proto-clusters at low and high redshift and with a considerable spatial resolution, allowing us to establish a firm relationship between their complex morphology and the mass assembly history. The recent exquisite observations operated by CFHT/Megacam, VLT/MUSE or ALMA already support the introduction of new spatial statistics besides the traditional ones calculated from the mass or inertia tensors (ellipticity, triaxiality, etc.), which are well-suited for relaxed or poorly resolved systems but less appropriate to describe merging clusters or their filamentary environment.

The usual morphological parameters can be impractical for diverse samples. Axial ratios and inertia eigenvectors provide a very accurate and physically motivated description of regular and approximately triaxial haloes, but they can fail to properly describe major mergers or bridges and filaments. In some sense, classic morphological schemes usually adopted in cluster astronomy can be properly used only after the shape of the halo as been already assessed. One first determines the class which the cluster belongs to and then adopts the relevant shape classifier.

The shapefinders based on the Minkowski functionals, introduced by \citet{SahniSathyaprakashShandarin1998} to investigate the morphology of the large scale structure and dubbed thickness ($H_1$), width ($H_2$), and length ($H_3$), provide instead a small set of parameters that can properly describe very diverse morphologies, possibly correlating with the entire accretion history of the halo. This study assesses the capability of Minkowski functionals and shapefinders to discriminate between ellipsoidal, merging, spiky, and dumbbell morphologies, providing explicit formulae for simplified geometries.

Equations (\ref{eq:ellipsoid}) for triaxial ellipsoids $\mathcal{E}$ and (\ref{eq:mergingspheres}) for two-fused balls $\mathcal{M}$ are the main analytical result of this study; to our knowledge the formulas for their integrated mean curvature, $H_\mathcal{E}$ and $H_\mathcal{M}$, are new in the literature. Using the additivity of Minkowski functionals and the Steiner formula, one can generalise the model to $n$ merging balls (satellites), $\mathcal{M}_n$. Analytical expression for axially-symmetric filaments with varying thickness, Equations~(\ref{eq:dumbbell1}-\ref{eq:dumbbell3}), are pivotal to build spiky geometries $\mathcal{S}_n$ accounting for filaments feeding a central halo or cluster-pair bridges $\mathcal{D}$.

It is important to remind that the (scalar) Minkowski functionals for the merger and spiky models, $\mathcal{M}_n$ and $\mathcal{S}_n$, in which the different satellites or branches do not mutually overlap, do not supply any information about the relative orientation of the substructures. The morphology of anisotropic bodies can be instead distinguished by the vector and tensor-valued Minkowski functionals \citep[e.g.][]{BeisbartValdarniniBuchert2001,Beisbart+2002}, which can be interpreted as generalisation of the moment of inertia of the body. Consistently, the so-called planarity and filamentarity shapefinders deduced from $(H_1,H_2,H_3)$ would be misleading for the simplified models considered here, thus not used for the classification.

Not surprisingly, as shown in Figure~\ref{fig:MFmergers+spiky}, the Minkowski functionals respond to the distance and size of satellites, to the connectivity of central haloes, and to the thickness of feeding filaments. These geometrical and topological properties trace the growth of structures and have an impact on the physical properties of galaxies in the nodes of the cosmic-web \citep{Choi+2010,CodisPogosyanPichon2018,Kraljic+2018,Kraljic+2020}. Reasonably enough, Minkowski functionals and shapefinders are therefore correlated with the mass assembly history both in the dark and gaseous components and could serve as diagnostics to investigate the relationship between local morphology and global dynamics.

Figures~\ref{fig:H123} and~\ref{fig:H123data} summarise this study. They show how a simple three-dimensional parameter space is adequate to describe the full variety of cluster. Thought not fully lifting the degeneracies necessarily occurring between very different morphologies, $(H_1,H_2,H_3)$ can be effectively used as classifiers provided at least some effective radius of the major structure is estimated.

This study is a proof-of-concept to illustrate the potential of Minkowski functionals and shapefinders for clusters studies. The full practical potential of this approach in cluster morphological analysis has still to be assessed. The toy models we considered can capture some of the main features of a diverse sample of clusters but likely fail to describe more complex configurations that shows up in the observed sky or in numerical simulations. In these case, alternative computation techniques shall be used. Depending on the discrete or continuous nature of the mass tracers, the underlying Minkowski functionals can be estimated using the so-called germ-grain or excursion set models \citep{MeckeBuchertWagner1994,Schmalzing+1999}, i.e. by dressing the point-processes  (e.g. galaxies or subhaloes) with balls of fixed radius, whose union forms the continuous body, or considering the iso-contours of suitably smoothed random field (e.g. the density or temperature of the cluster), respectively using the radius of balls or the threshold value defining the iso-contours as diagnostic parameter.

We showed the potential of shapefinders as morphological classifier in 3D. The three-dimensional shape of galaxy clusters can be constrained with joint multi-wavelength analyses combining lensing, X-ray, and SZ \citep{Sereno+2018} or deep spectroscopic campaigns \citep{Rosati+2014,Finoguenov2019,Kuchner+2020}, which unveil the third dimension orthogonal to the projected sky. The data sets required by these analyses can be very expansive and the full 3D analysis of galaxy clusters is usually not feasible for most of the known halos. Even though this situation can change with the next generation surveys and instruments, it might be useful to consider 2D shapefinder classes in the projected space. This could be more useful in the context of large surveys.

Finally, it is worth to stress that morphology alone cannot unambiguously determine the degree of equilibrium of a halo. Apparently, morphological regular clusters can be unrelaxed \citep{Meneghetti+2014}. Any possible correlation between shapefinders and dynamical state of the clusters shall require a more accurate investigation, also based on $N$-body simulations.

\vspace{-12pt}
\section*{Acknowledgments}
We thank G. Covone, K. Kraljic, and E. Sarpa for discussions and a critical reading of the manuscript, and the anonymous referee for the fruitful suggestions that largely improved the illustration of our results. This work has been partially supported by the Programme National Cosmology et Galaxies (PNCG) of CNRS/INSU with INP and IN2P3, co-funded by CEA and CNES, and Labex OCEVU (ANR-11-LABX-0060). MS acknowledges financial contribution from contracts ASI-INAF n.2017-14-H.0 and INAF mainstream project 1.05.01.86.10. MS acknowledges LAM for hospitality.

\renewcommand{\thefigure}{A\arabic{figure}}
\setcounter{figure}{0}

\begin{figure*}
\begin{center}
\includegraphics[width=\textwidth]{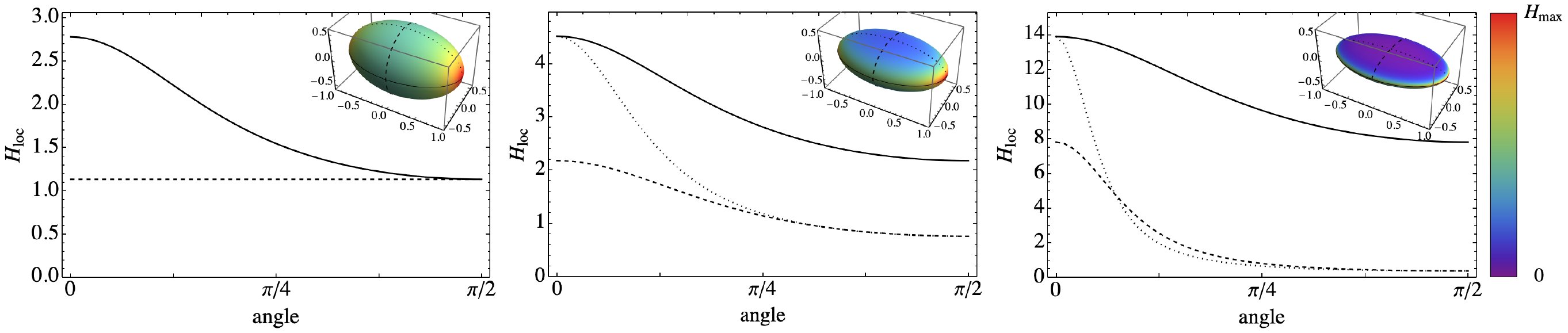}
\caption{Local mean curvature $H_\mathrm{loc}$ of triaxial ellipsoids (inset) with axes ratio $(b/a,c/a)=(0.6,0.6)$ (left), $(0.6,0.4)$ (centre), and  $(0.6,0.2)$ (right) as function of the angles measured from the centre of the body and spanning a quarter of the equatorial plane (solid line) and of the two perpendicular meridian planes (dashed, dotted; same ). For the axially symmetric, prolate ellipsoid (left) two directions are equivalent. Note the different range of values of $H_\mathrm{loc}$.}
\label{appendix:fig:ellipsoid:localcurvature3dCH}
\end{center}
\end{figure*}

\section*{Data Availability}

The data underlying this article are available in the article and in its online supplementary material.



\bibliographystyle{mnras}
\bibliography{MergingClusters_MinkowskiFunctionals} 


\appendix

\section{Integrated mean curvature of an ellipsoid}\label{sec:appendix:ellipsoid}

Following \cite{Poelaert+2011arXiv1104.5145P}, an ellipsoid described by
\begin{equation}
\frac{X^2}{a^2}+\frac{Y^2}{b^2}+\frac{Z^2}{c^2}=1,
\end{equation}
with principal semi-axes $a\geqslant b\geqslant c$ and central Cartesian coordinates
\begin{align}
X &= a \cos\theta, \\
Y &= b \sin\theta\cos\phi, \\
Z &= c \sin\theta\sin\phi,
\end{align}
written 
in terms of the eccentric anomalies $\theta$ and $\phi$, has local mean curvature
\begin{equation}
H_\mathrm{loc}(\theta,\phi)=\frac{h^3(a^2+b^2+c^2-R^2)}{2a^2b^2c^2},
\end{equation}
with
\begin{equation}
h=\frac{abc}{\sqrt{b^2c^2\cos^2\theta+a^2(c^2\cos^2\phi+ b^2\sin^2\phi)\sin^2\theta}}
\end{equation}
being the shortest distance (`height') from the centre to the tangent plane to the ellipsoid at the point considered and $R=\sqrt{X^2+Y^2+Z^2}$ the radius to this point upon the ellipsoid surface (see Figure~\ref{appendix:fig:ellipsoid:localcurvature3dCH}). The local Gaussian curvature is $G_\mathrm{loc}=h^4/a^2b^2c^2$.

The mean curvature integrated over the surface is
\begin{equation}\label{eq:ellipsoid:curvature}
H \equiv abc \int_0^{2\pi}\dd\phi \int_0^\pi \dd\theta\,\frac{\sin\theta}{h}H_\mathrm{loc} = \frac{abc}{a^2} \,(I_1+I_2),
\end{equation}
with
\begin{subequations}\label{eq:ellipsoid:curvature:Ii}
\begin{align}
I_1 &= \int_0^{2\pi}\frac{a^2+k^2}{k^2} \frac{\mathrm{arctanh}~U}{U}\dd\phi, \\
I_2 &= \int_0^{2\pi}\frac{a^2-b^2-c^2-k^2}{k^2} \left(\frac{1}{U^2}-\frac{\mathrm{arctanh}~U}{U^3}\right)\dd\phi,
\end{align}
\end{subequations}
where $U^2=1-\frac{b^2 c^2}{a^2 k^2}\leq 1$ and $k^2=b^2\sin^2\phi+c^2\cos^2\phi$. The dimensionless integrals (\ref{eq:ellipsoid:curvature:Ii}) are evaluated numerically. Analytic limits exists for prolate and oblate ellipsoids (see main text).


\section{Integrated mean curvature of two merged balls}\label{sec:appendix:mergingspheres}

\begin{figure}
\begin{center}
\includegraphics[height=5.7cm]{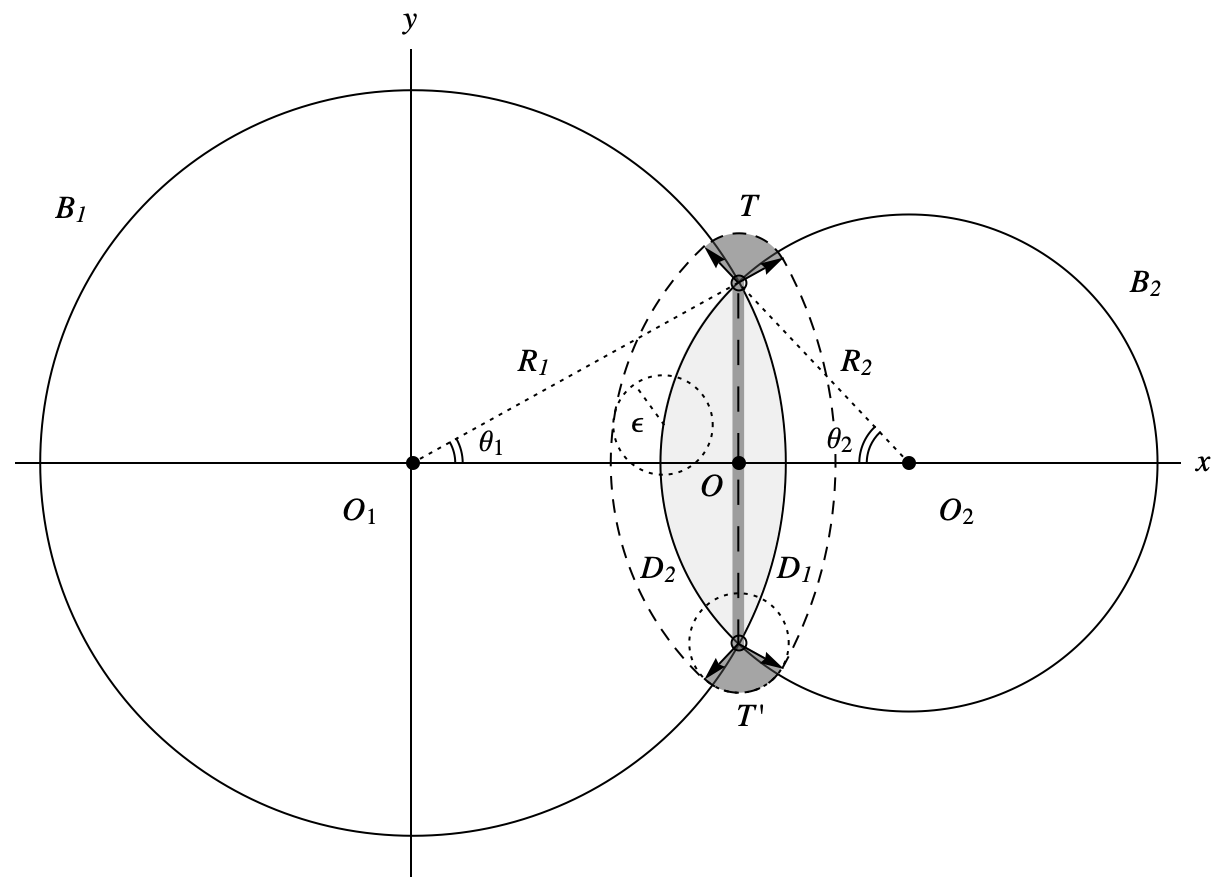}
\caption{Section of two fused balls $B_1$ and $B_2$ with centres separated by a distance $r$ and radius $R_1$ and $R_2$. Their intersection $\mathcal{L}\equiv B_1\cap B_2$ forms a ``bi-concave lens'' (light grey). Covering $\mathcal{L}$ with balls with radius~$\epsilon$ (only two are shown, dotted) one obtains the parallel lens $\mathcal{L}_\epsilon$ (dashed), which results in the union of two axially-symmetric dihedra $D_1$ and $D_2$ glued to a wedged torus (sections $T$ and $T'$ in dark gray).}
\label{appendix:fig:mergingballs}
\end{center}
\end{figure}

The integrated mean curvature of $\mathcal{M}=B_1\cup B_2$ is calculated using $V_2(B_1\cup B_2)=V_2(B_1)+V_2(B_2)-V_2(B_1\cap B_2)$. The last term is derived from the Steiner formula applied to parallel lens $\mathcal{L}_\epsilon$, namely the uniform coverage of the lens $\mathcal{L}\equiv B_1\cap B_2$ by balls with radius~$\epsilon$. As illustrated in Figure~\ref{appendix:fig:mergingballs}, $\mathcal{L}_\epsilon$ results in the union of $\mathcal{L}$ with two dihedra $D_1$ and $D_2$ of thickness $\epsilon$ and opening angles respectively $\theta_1$ and $\theta_2$, and a wedged torus $T$ centred on $O$ and with cross-section $(\pi-\theta_1-\theta_2)\epsilon^2$. According to the Steiner formula, its volume $V(\mathcal{L}_\epsilon)\equiv V(\mathcal{L})+V(D_1)+V(D_2)+V(T)$ is equal to
$V(\mathcal{L})+A(\mathcal{L})\epsilon+H(\mathcal{L})\epsilon^2+\frac{4\pi}{3}\epsilon^3$ \citep{Mecke2000}, i.e. a fourth-order polynomial in $\epsilon$ with coefficients proportional to the Minkowski functionals; the integrated mean curvature $H(\mathcal{L})$ corresponds to the sum of the terms of $V(\mathcal{L}_\epsilon)$ proportional to $\epsilon^2$.

The volume of the spherical dihedron $D_1$ is
\begin{equation}\label{eq:voldihedron}
V(D_1)=2\pi\left(1-\frac{p}{R_1}\right)\left(R_1^2\epsilon+R_1\epsilon^2+\frac{\epsilon^3}{3}\right),
\end{equation}
with $p\equiv\overline{OO}_1=(R_1^2-R_2^2+r^2)/2r$ the distance from the centre of $B_1$ and the centroid of the lens and $r\equiv\overline{O_1O}_2$. A similar expression holds for $V(D_2)$ replacing $R_1$ by $R_2$ and $p$ by $r-p$. The volume of the wedged torus $T$ is
\begin{equation}\label{eq:volwedgetorus}
V(T)=\pi(\pi-\theta_1-\theta_2)\rho\epsilon^2+\frac{2\pi}{3}(\cos\theta_1+\cos\theta_2)\epsilon^3,
\end{equation}
with $\rho=(R_1^2-p^2)^{1/2}$ its major radius, $\cos\theta_1=p/R_1$, and $\cos\theta_2=(r-p)/R_2$. The Steiner formula finally yields
\begin{align}
V(\mathcal{L})&=\frac{\pi}{12}r^3 - \frac{\pi}{4}\frac{\Delta^4}{r} - \frac{\pi}{2}r \Sigma^2 + \frac{2\pi}{3}(R_1^3 + R_2^3)\;, \\
A(\mathcal{L})&= 2\pi\Delta^2-\pi r(R_1+R_2)-\frac{\pi}{r}(R_1-R_2)\Delta^2\;, \\
H(\mathcal{L})&= 2\pi(R_1+R_2-r) + \pi\psi\sqrt{2\Sigma^2-r^2-\frac{\Delta^4}{r^2}}\;,
\end{align}
in which $\cos\psi\equiv\cos(\pi-\theta_1-\theta_2)=(\Sigma^2 - r^2)/(2R_1R_2)$, $\Sigma^2=R_1^2+R_2^2$, and $\Delta^2=R_1^2-R_2^2$. All these equations are valid for non trivial intersection, i.e. overlap with no embedding or $|r-R_1| \leqslant R_2$.

\section{Minkowski functionals of the multiple-mergers model \texorpdfstring{$\boldsymbol{\mathcal{M}^*_n}$}{Sstar}}\label{sec:appendix:MultiMergersModel}

The Minkowski functionals of $\mathcal{M}^*_n=B\cup B_1\cup\dots\cup B_n$ for $i=1,\dots,n$ satellites with radius $r_i<r$, not mutually overlapping, and avoiding trivial embedding ($|d_i-r|\leqslant r_i$) are
\begin{subequations}\label{eq:stellarspheres}
\begin{align}
V_0^{\mathcal{S}^*}&= \frac{2\pi}{3}(2-n)r^3+\frac{2\pi}{3}\sum_i\left(r_i^3-\frac{1}{8}d_i^3\right) \nonumber \\
	& \; + \frac{\pi}{2}\sum_i d_i(r^2+r_i^2)+ \frac{\pi}{4}\sum_i\frac{(r^2-r_i^2)^2}{d_i}, \label{eq:mergingspheresV0} \\
V_1^{\mathcal{S}^*}&= \frac{\pi}{3}(2-n) r^2 - \frac{\pi}{3}\sum_i r_i^2 \nonumber\\
	& + \frac{\pi}{6}\sum_i d_i(r+r_i) + \frac{\pi}{6}\sum_i \frac{(r-r_i)(r^2-r_i^2)}{d_i},\\
V_2^{\mathcal{S}^*}&= \frac{2}{3}r + \frac{2}{3}\sum_i(r_i+d_i) \nonumber \\ 
	& - \frac{1}{3}\sum_i \psi_id_i\sqrt{2\frac{r^2+r_i^2}{d_i}-1-\left(\frac{r^2-r_i^2}{d_i^2}\right)^2},
\end{align}
\end{subequations}
with $\cos\psi_i=(r^2+r_i^2-d_i^2)/2rr_i$. The condition $B_i\cap B_j = \emptyset$ is approximately realised if the surface covered by the basis of the $n$ balls does not exceed the surface of the central sphere, $\sum_i [(r^2-r_i^2+d_i^2)/(2d_i)]^2\lesssim 4r^2$.


\section{Minkowski functionals of the dumbbell model \texorpdfstring{$\boldsymbol{\mathcal{D}_P}$}{Dumbb}}\label{sec:appendix:dumbbell}

\begin{figure}
\begin{center}
\includegraphics[width=8.3cm]{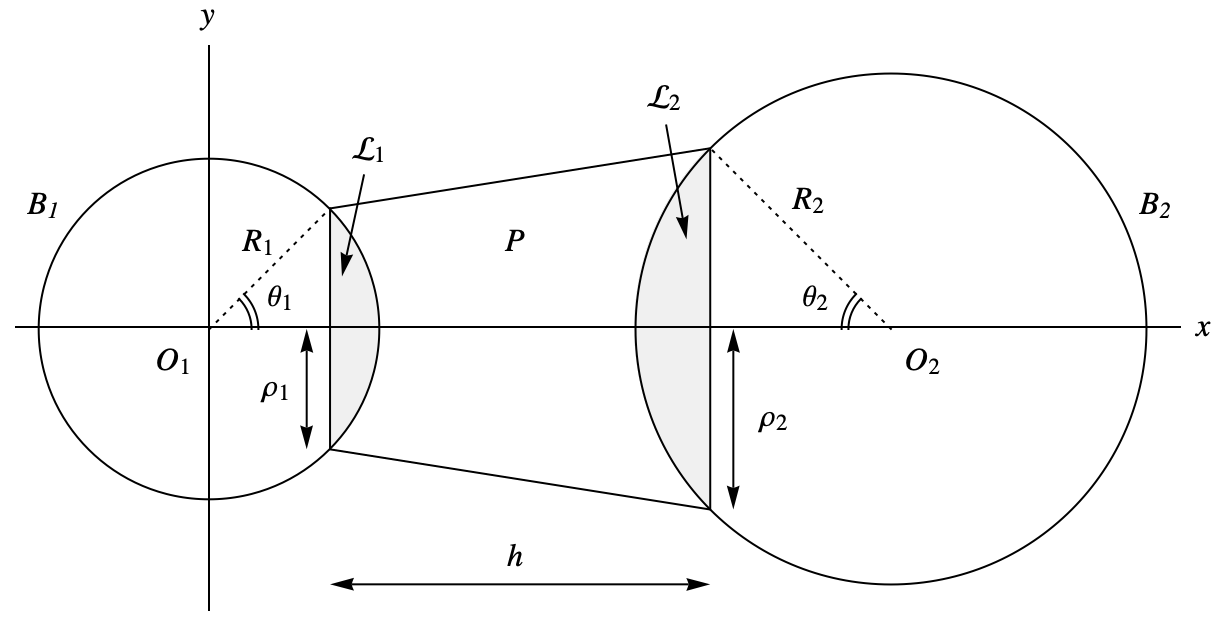}
\caption{Section of dumbbell model with axially-symmetric conic filament $P$. The intersection of the two balls $B_1$ and $B_2$ with $P$ yields two ``flat-concave lenses'' $\mathcal{L}_1$ and $\mathcal{L}_2$ (light grey).}
\label{appendix:fig:dumbbell}
\end{center}
\end{figure}

The dumbbell $\mathcal{D}_P=B_1\cup P \cup B_2$ resulting from the union of two balls $B_i$ ($i=1,2$) bridged by a truncated cone $P$ (Figure~\ref{appendix:fig:dumbbell}) has $V_\mu(\mathcal{D}_P)=V_\mu(B_1)+V_\mu(B_2)+V_\mu(P)-V_\mu(\mathcal{L}_1)-V_\mu(\mathcal{L}_2)$. The Minkowski functionals of $\mathcal{L}_i\equiv B_i\cap P$ and $P$ are calculated as proportional to $\epsilon^\mu$ (Steiner formula) using Equations~(\ref{eq:voldihedron},\ref{eq:volwedgetorus}). The non-trivial result for $P$ are
\begin{align}
V(P)&= \pi\rho_1^2h-\pi\rho_1h^2\tan\theta+\frac{\pi}{3} h^3\tan^2\theta,\label{eq:dumbbell1}
\\
A(P)&= \pi(\rho_1^2+\rho_2^2) +2\pi h(\rho_1\cos\theta+h\sin\theta) \label{eq:dumbbell2}
 \\
	& +\pi(\rho_2^2-\rho_1^2)\sin\theta, \nonumber \\
H(P)&= \pi h\cos^2\theta +\pi(\pi-2\theta)\frac{\rho_1+\rho_2}{2} -\pi\frac{\rho_2-\rho_1}{2}\sin2\theta, 
	\label{eq:dumbbell3}
\end{align}
where $\rho_1$ and $\rho_2$ are the radii of the minor and major circular basis of $P$, $h$ its height, and $\tan\theta=(\rho_2-\rho_1)/h$. For $\theta=0$ one recovers the known Minkowski functionals for a cylinder $C$ with basis $\rho\equiv\rho_1=\rho_2$ and height $h$, i.e. $V(C)=\pi\rho^2h$, $A(C)=2\pi\rho(\rho+h)$, $H(C)=\pi(h+\pi\rho)$. For $h=0$ the second and third Minkowski functionals further yield the area and the integrated mean curvature (or $2/\pi~\times$ perimeter) of a two-faces two-dimensional disk embedded in a three-dimensional space.


\bsp
\label{lastpage}

\end{document}